# Broadband Photoresponse Arising from Photo-Bolometric Effect in Quasi-One-Dimensional $Ta_2Ni_3Se_8$

W L Zhen[1,2], W T Miao[1], W L Zhu[1], C J Zhang[1,3,*] and W K Zhu[1,*]

[1] High Magnetic Field Laboratory, Chinese Academy of Sciences, Hefei 230031, People's Republic of China

[2] University of Science and Technology of China, Hefei 230026, People's Republic of China

[3] Institutes of Physical Science and Information Technology, Anhui University, Hefei 230601, People's Republic of China

E-mail: zhangcj@hmfl.ac.cn and wkzhu@hmfl.ac.cn

## Abstract

In this paper, we report the synthesis of high-quality $Ta_2Ni_3Se_8$ crystals free of noble or toxic elements and the fabrication and testing of photodetectors on the wire samples. A broadband photoresponse from 405 nm to 1550 nm is observed, along with performance parameters including relatively high photoresponsivity (10 mA $W^{-1}$) and specific detectivity ($3.5 \times 10^7$ Jones) and comparably short response time ($\tau_{rise}$ = 433 ms, $\tau_{decay}$ = 372 ms) for 1064 nm, 0.5 V bias and 1.352 mW $mm^{-2}$. Through extensive measurement and analysis, it is determined that the dominant mechanism for photocurrent generation is the photo-bolometric effect, which is believed to be responsible for the very broad spectral detection capability. More importantly, the pronounced response to 1310 nm and 1550 nm wavelengths manifests its promising applications in optical communications. Considering the quasi-one-dimensional structure with layered texture, the potential to build nanodevices on $Ta_2Ni_3Se_8$ makes it even more important in future electronic and optoelectronic applications.

## 1. Introduction

Photodetectors, with the ability to convert incident light into electrical signals, are essential for industrial and scientific applications [1-3]. Generally, photodetectors can be divided into two types, i.e., specific wavelength detectors (ultraviolet (UV) detectors, visible light (vis) detectors or infrared (IR) detectors) and broadband detectors (UV-IR detectors) [4], according to the detection spectral range. Specific wavelength photodetectors need to work under a certain single wavelength or a narrow range of light, showing specific applications in light detection and nanophotonic integrated circuits [5, 6]. In contrast, broadband photodetectors have enormous potential applications in UV−vis−IR light communications, memory storage and broad-spectrum switching in a single optoelectronic system [1, 7, 8]. In the past few years, many commercial photodetectors are typically based on crystalline GaN [9], Si [10] and InGaAs [11] and used for 250−400 nm (UV), 400−800 nm (vis) and 900−1700 nm (near-IR) detection, respectively. In addition, longwave IR detectors based on narrow bandgap or semimetal materials have been extensively studied, such as HgCdTe [12], InSb [13, 14] and $Cd_3As_2$ [15]. Although these materials show good performance in the near-IR and longwave IR, photodetectors based on these materials contain toxic elements and often require low temperatures to achieve high performance, which makes the systems bulky, expensive, and harmful to humans and the environment [16]. Therefore, there is a great need to develop high-performance IR photodetectors that operate at room temperature and consist only of low-cost and nontoxic elements.

Recently, the quasi-one-dimensional (quasi-1D) chain compounds $M_2X_3Y_8$ (M= Ta, Nb; X= Ni, Pd, Pt; Y=S, Se) have drawn considerable attention for their intriguing electronic properties related to the quasi-1D structure [17-20]. For the crystal structure, the atoms in the chains are connected by strong chemical bonds, while the inter-chain bonds are relatively weak, i.e., the van-der-Waals (vdW) interactions. Crystals of this type usually show a needle-like or fine hair morphology. For the electronic structure, the $Ta_2Pd_3Se_8$ and $Ta_2Pt_3Se_8$ transistors are n-type and p-type semiconductors, respectively [18]. Meanwhile, these materials also exhibit outstanding optoelectronic properties [21, 22]. Among these compounds, $Ta_2Ni_3Se_8$ is a low-cost material free of noble or toxic elements, unlike its analogues with noble metals such as Pd and Pt. $Ta_2Ni_3Se_8$ nanowire transistor has been identified as an ambipolar semiconductor [20]. Considering the interesting electronic structure and properties, $Ta_2Ni_3Se_8$ holds promise for possible applications in the field of photodetection, but it remains to be explored.

In this work, we report the synthesis of high-quality $Ta_2Ni_3Se_8$ crystals and the fabrication and testing of photodetectors on the wire samples. A broadband photoresponse from 405 nm to 1550 nm is observed, along with performance parameters including relatively high photoresponsivity (10 mA $W^{-1}$) and specific detectivity ($3.5 \times 10^7$ Jones) and comparably short response time ($\tau_{\mathrm{rise}}$ = 433 ms, $\tau_{\mathrm{decay}}$ = 372 ms) for 1064 nm, 0.5 V bias and 1.352 mW $mm^{-2}$. Through extensive measurement and analysis, it is determined that the dominant mechanism for photocurrent generation is the photo-bolometric (PB) effect, which is believed to be responsible for the very broad spectral detection capability. More importantly, the pronounced response to 1310 nm and 1550 nm wavelengths manifests its promising applications in optical communications. The quasi-1D structure with layered texture and the potential to build nanodevices make it even more important.

## 2. Methods

$Ta_2Ni_3Se_8$ crystals were prepared by a one-step method [20]. High purity powders of Ta (99.98%, Alfa Aesar), Ni (99.9%, Alfa Aesar) and Se (99.9%, Alfa Aesar) were mixed and ground in a chemical molar ratio of 2:3:10. The mixture was then sealed in a 12 cm long evacuated quartz ampule, heated to 700 °C at a rate of 2 °C/min, and kept at this temperature for 7 days before cooling down to room temperature. After the reaction, large needle-like or fine hair-like $Ta_2Ni_3Se_8$ crystals can be obtained on the surface of the solid solution. See the supplementary information for the optical image. The crystal structure and phase purity were checked by X-ray diffraction (XRD) analysis on a Rigaku-TTR3 X-ray diffractometer using Cu K$\alpha$ radiation at room temperature. Chemical composition characterization was performed on an Oxford Inca energy dispersive spectrometer (EDS) equipped with a FEI Helios Nanolab 600i scanning electron microscope (SEM). Raman spectrum was collected in a backscattering geometry on a Horiba JobinYvon T6400 spectrometer with a 532 nm laser.

Photodetectors were fabricated based on $Ta_2Ni_3Se_8$ wires. Gold wires were directly connected to the sample using silver paste for good ohmic contact. According to the previous reference [20], the use of chromium as the contact electrodes resulted in a relatively low Schottky barrier of 50 meV, which was a nearly ohmic contact. Considering the low work function of chromium (4.5 eV) and the ambipolar semiconductor characteristic of $Ta_2Ni_3Se_8$, we adopted another low work function metal, i.e., silver (4.3 eV). The channel area of the device on $SiO_2$ (300 nm)/Si substrate was about 1 μm × 2 mm, and the thickness of the active material was about 2 μm. The optoelectronic measurements were carried out on a home-built multi measurement system consisting of a microscope, a four-probe station, a semiconductor parameter analyzer (2636B, Keithley) and laser light sources with different wavelengths. The incident light power density was measured using a power meter (PM100D, Thorlabs). The incident laser spot diameter was approximately 5 mm, which is larger than the channel area to ensure uniform illumination. The electronic structure of bulk $Ta_2Ni_3Se_8$ was theoretically investigated by first-principles calculations using the WIEN2K code [23]. The generalized gradient approximation (GGA) of Perdew-Burke-Ernzerhof [24] was employed with a $k$-mesh of 1000 points. The lattice constants used for the calculations were taken from the Springer Materials website.

## 3. Results and discussion

The crystal structure of $Ta_2Ni_3Se_8$ is orthorhombic with the space group *Pbam*. It is found that the unit cell of $Ta_2Ni_3Se_8$ contains four interconnected chains forming a zigzag structure extending along the $c$-axis. Each chain consists of two edge-sharing chains of trigonal prisms centered on Ta atoms and bridged by Ni atoms, as shown in figure 1(a).

Between linear chains are vdW gaps, which facilitate the quasi-1D structure. Figures 1(b) and (c) show the band structure and density of states (DOS) of $Ta_2Ni_3Se_8$ along the high symmetry directions, respectively. The calculated bandgap is about 0.25 eV, indicating that it is a narrow bandgap semiconductor. The obtained value is close to the reported results [20], confirming the reliability of our calculations. From the band structure, the bandgap is an indirect one since the valence band maximum (VBM) and conduction band minimum (CBM) do not coincide at the same point. The VBM and CBM appear to be in the UX and YZ segments, respectively. From the DOS plots, the CBM and VBM are mainly composed of distorted *d* and *p* orbitals of Ta, Ni and Se ions, revealing the nature of covalent bonding.

The obtained crystals were characterized by XRD, EDS and Raman spectroscopy. Figure 2(a) shows the XRD pattern of a bunch of thin wires. All diffraction peaks can be indexed, which is in good agreement with the powder XRD results (JCPDS PDF No. 00-086-0186), indicating that the compound is single-phase $Ta_2Ni_3Se_8$. Figure 2(b) shows a typical SEM image of a $Ta_2Ni_3Se_8$ wire. The surface of the wire is flat and smooth. Interestingly, the side view reveals a feature of layered texture, suggesting the possibility of exfoliation. The EDS spectrum collected on the wire surface indicates a chemical composition of Ta:Ni:Se = 16.44%:23.62%:59.94% (figure 2(c)), which is very close to the stoichiometric ratio of 2:3:8. Note that due to the low melting point (221°C) and high saturation vapor pressure of Se, there is a certain degree of Se vacancies. Furthermore, figure 2(d) shows the EDS mapping results of the total scan and elemental scan of Ta, Ni and Se atoms, respectively. The uniform distribution of the elements confirms the high quality of the crystals. More EDS data are presented in the supplementary information. In addition, the Raman spectrum shown in the supplementary information is consistent with a recent report [25]. No photoluminescence signal was observed due to the indirect bandgap feature.

To investigate the optoelectronic properties of $Ta_2Ni_3Se_8$, several two-terminal devices with similar dimensions were fabricated based on $Ta_2Ni_3Se_8$ wires and tested. Since there are no significant differences between these devices, we only provide experimental results for a representative device here (see the supplementary information for other devices). In figure 3, we summarize the optoelectronic measurements of device D1 in the dark and under laser illumination of different wavelengths ($\lambda$), i.e., 405 nm, 450 nm, 488 nm, 635 nm, 780 nm, 830 nm, 940 nm and 1064 nm. The laser power density is tuned to a constant value of 1.352 mW mm$^{-2}$. As shown in figure 3(a), the current-voltage (*I*-*V*) characteristic curves pass through the origin and are almost completely linear with ±0.5 V, indicating a good ohmic contact between $Ta_2Ni_3Se_8$ and the electrodes, which also rules out the possibility of Schottky contact. The inset shows an enlarged view of the *I*-*V* curves in the range of 0.48 V to 0.5 V. The dark current at 0.5 V corresponds to a current density of 256 A cm$^{-2}$. A clear difference between the dark and illuminated states is observed at all laser wavelengths, implying that the photodetector responds to a broad spectral range from UVA to the entire visible spectrum to the near-IR. Figure 3(b) shows the calculated photocurrents for various wavelengths under 0.5 V bias. The photocurrent is defined as $I_{\mathrm{ph}} = I_{\mathrm{illumination}} - I_{\mathrm{dark}}$, where $I_{\mathrm{illumination}}$ and $I_{\mathrm{dark}}$ are the currents measured in the illuminated and dark states, respectively. From figure 3(b), the photocurrent at $\lambda$ = 405 nm is remarkable, reaching 120 nA. Although it degrades for longer wavelengths, the photocurrent in the IR region is significant enough. Given that the calculated bandgap of $Ta_2Ni_3Se_8$ is 0.25 V, which corresponds to a wavelength of 4959 nm, the wavelengths used in the present work far exceed the bandgap. Shorter wavelength photons have higher energy and will excite more electrons deep in the valence band [26]. This could explain the phenomenon that larger photocurrents appear at shorter wavelength. This is a preliminary judgement. In-depth studies on the mechanism will be presented in the next section.

To further evaluate the optoelectronic performance, some key parameters are calculated, including responsivity ($R = \frac{I_{\mathrm{ph}}}{PS}$, where $P$ is the laser power density and $S$ is the illuminated area) representing the conversion efficiency from incident light signal to electrical signal, specific detectivity ($D^* = R\sqrt{\frac{S}{2eI_{\mathrm{dark}}}}$, where $e$ is the elementary charge) characterizing the capability to detect the lowest light signal, and external quantum efficiency (EQE = $\frac{hc}{e\lambda}R \times 100\%$, where $c$ is the speed of light in vacuum) referring to the number of electrons excited by each incident photon [27, 28]. Figures 3(c) and (d) show $R$ (left axis), $D^*$ (right) and EQE as a function of wavelength, respectively. The most significant parameters are recorded as $R$ = 43.3 mA $W^{-1}$, $D^*$ = 1.5 × $10^8$ Jones and EQE = 13.3%, appearing at $\lambda$ = 405 nm and $P$ = 1.352 mW $mm^{-2}$. Meanwhile, when the photodetector is exposed to the 1064 nm near-IR light with the same power density, relatively high responsivity ($R$ = 10 mA $W^{-1}$) and specific detectivity ($D^*$ = 3.5 × $10^7$ Jones) can be obtained. These parameters are significant, even in comparison with other outstanding photodetectors fabricated on ternary compounds [29-34] or quasi-1D wires [29, 30, 35, 36]. Considering the broadband photoresponse, the advantage of $Ta_2Ni_3Se_8$ photodetector is more prominent.

The dependence of optoelectronic properties on laser power density is further studied. Considering the harmless feature of visible light, the 635 nm laser is selected to perform more measurements. As shown in figure 4(a), the *I-V* curves exhibit a clear dependence on power density. The photocurrent is calculated and plotted as a function of bias voltage for different power densities ranging from 0.16 to 1.352 mW $mm^{-2}$ (figure 4(b)). Two characteristics can be observed. First, the photocurrent increases as the power density increases. Second, all the curves are near linear, which is easy to understand because of the linear dependence of the curves in figure 4(a). The inset of figure 4(b) shows the time-resolved photocurrent taken under different power densities and 0.01 V bias. The ladder-like curve shows well-defined ON and OFF states, indicating good stability and nonvolatile feature of the device. The photocurrent is about 9 nA at 1.352 mW $mm^{-2}$ and decreases to 1 nA at 0.16 mW $mm^{-2}$. Upon further decreasing the power density, no obvious photocurrent is detected. Additional experiments performed to test the long-term stability can be found in the supplementary information.

The relationship between photocurrent and power density is important for studying the underlying mechanism [7, 30, 31]. Generally, the photocurrent is a power function of the power density, i.e., $I_{ph} \propto P^{\alpha}$, where $\alpha$ is the power exponent. It is suggested that $\alpha = 1$ corresponds to the ideal photoconductive (PC) effect and $\alpha < 1$ indicates existence of photogating effect arising from trapping states [31]. Figure 4(c) shows the photocurrent plotted as a function of power density at 0.1 V, 0.3 V and 0.5 V biases. The data points are fitted to the power law, which give $\alpha$ values of 1.14, 1.16 and 1.14, respectively. These values appear to be independent of bias, which is consistent with the linear photocurrent-bias relationship mentioned above (figure 4(b)). The consistent values also confirm the reliability of the fits. Unexpectedly, $\alpha$ here is larger than 1, which suggests another mechanism than the photoconductive or photogating effect. Usually, $\alpha$ over 1 is related to photo-bolometric effect [37, 38]. Since the laser spot covers the entire channel area, the channel of the photodetector is heated evenly. That is, thermal gradients are negligible and thermoelectric mechanism can be ruled out. The generation of photocurrent is simply attributed to the decrease in channel resistance caused by heating due to photon absorption (see figure 5(a) for a schematic illustration of this effect). In addition to the power-law fits, more evidence can support the determination of the photo-bolometric effect. First, this effect requires, in principle, that the photocurrent is linearly proportional to the bias voltage. This is consistent with our measurements (figure 4(b)). Second, the photo-bolometric effect leads to a gradual increase in the responsivity and detectivity [37]. This can also be seen in our case, as shown in figure 4(d). The responsivity here can be formulated as $R = \frac{I_{ph}}{PS} \propto \frac{P^{\alpha}}{PS} \propto P^{0.14}$. It is thus reasonable that the responsivity increases with increasing power density and gradually approaches saturation. Third, this effect does not involve the bandgap width [7]. Therefore, a broadband photoresponse is expected. This is exactly the most important feature of $Ta_2Ni_3Se_8$ photodetectors. All these facts are self-consistent and demonstrate the dominant role of the photo-bolometric effect.

Furthermore, the response time of the $Ta_2Ni_3Se_8$ photodetector is studied in detail. Figures 5(b)-(i) show the time-resolved photocurrent taken at 0.01 V bias and $P$ = 1.352 mW $mm^{-2}$ under periodic illumination of different wavelengths (405-1064 nm). All curves exhibit good switching cycle stability, indicating that the photodetector can operate well and persistently at various wavelengths. As a commonly used method, the rising time ($\tau_{rise}$) is defined as the time for the photocurrent to increase from 10% to 90% of the maximum value, and the recovery time ($\tau_{decay}$) is the time for the photocurrent to decrease from 90% to 10%. As shown in the insets of figures 5(b)-(i), the determination process gives relatively short response time ($\tau_{rise}$ < 600 ms and $\tau_{decay}$ < 820 ms) in the detection spectral range, which is comparable to the performance of some benchmark photodetectors reported previously, such as monolayer $MoS_2$ [39], $Ta_2Se_8I$ [37], SnSe [40] and graphene oxide [41]. The fastest response occurs under 830 nm illumination, i.e., $\tau_{rise}$ = 360 ms and $\tau_{decay}$ = 400 ms. In addition, the good square waveforms indicate that the devices are not RC constant limited for the given parameters. See the supplementary information for the relationship between the RC constant and the photocurrent-time waveform and the power density dependence of the response time.

Given the prominent performance and the broadband photoresponse, it is natural to extend the testing further into longer wavelength region, such as the telecommunication band, which is important for today's industrial and consumer electronics [3, 42]. Here we select two typical wavelengths in the telecommunication band, namely O band 1310 nm and C band 1550 nm for testing. Figure 6(a) presents a schematic illustration of the $Ta_2Ni_3Se_8$ photodetector under global telecommunication wavelength illumination. As shown in figure 6(b), linear *I-V* curves together with clear photoresponse are again observed. Figures 6(c) and (d) show the time-resolved photocurrent at 0.01 V bias under periodic radiation of 1310 nm and 1550 nm, respectively. The device again exhibits good stability and durability. In the meanwhile, the response is also fast, represented by short response time, namely $\tau_{rise}$ = 430 ms and $\tau_{decay}$ = 414 ms for 1310 nm and $\tau_{rise}$ = 354 ms and $\tau_{decay}$ = 345 ms for 1550 nm. These values are shorter than the response time of other IR materials, including $Ta_2NiSe_5$ [43], $VO_2$ [44] and $Ta_2Se_8I$ [37]. This result is of great significance because the pronounced response to 1310 nm and 1550 nm suggests promising applications in optical communications and fiber optic cable testing in O band and C band. To further highlight the performance of $Ta_2Ni_3Se_8$ as a broadband photodetector, a detailed comparison with other reported materials is listed in Table 1. Considering the quasi-1D layered feature in its structure, the potential to build nanodevices on $Ta_2Ni_3Se_8$ makes it even more important in future electronic and optoelectronic applications.

## 4. Conclusion

In conclusion, we have synthesized high-quality $Ta_2Ni_3Se_8$ crystals and fabricated photodetectors on the wire samples. The photodetector exhibits relatively high photoresponsivity (10 mA $W^{-1}$) and specific detectivity ($D^* = 3.5 \times 10^7$ Jones) and comparably short response time ($\tau_{rise} = 433$ ms, $\tau_{decay} = 372$ ms) for $\lambda = 1064$ nm, 0.5 V bias and $P = 1.352$ mW $mm^{-2}$. Through extensive measurement and analysis, it is determined that the dominant mechanism for photocurrent generation is the photo-bolometric effect. This effect is believed to be responsible for the very broad spectral detection capability, extending into the mid- to long-IR region. More importantly, the pronounced response to 1310 nm and 1550 nm wavelengths manifests its promising applications in optical communications. Future work can focus on the quantum confinement effects at the nanoscale, such as optimizing the optoelectronic properties by manufacturing nanodevices and applying gating techniques.

## Acknowledgments

This work was supported by the National Key R&D Program of China (Grant No. 2021YFA1600201), the National Natural Science Foundation of China (Grant Nos. 11874363, 11974356 and U1932216) and Anhui Province Laboratory of High Magnetic Field (Grant No. AHHM-FX-2020-01).

## Data availability statement

All data that support the findings of this study are included within the article (and any supplementary files).

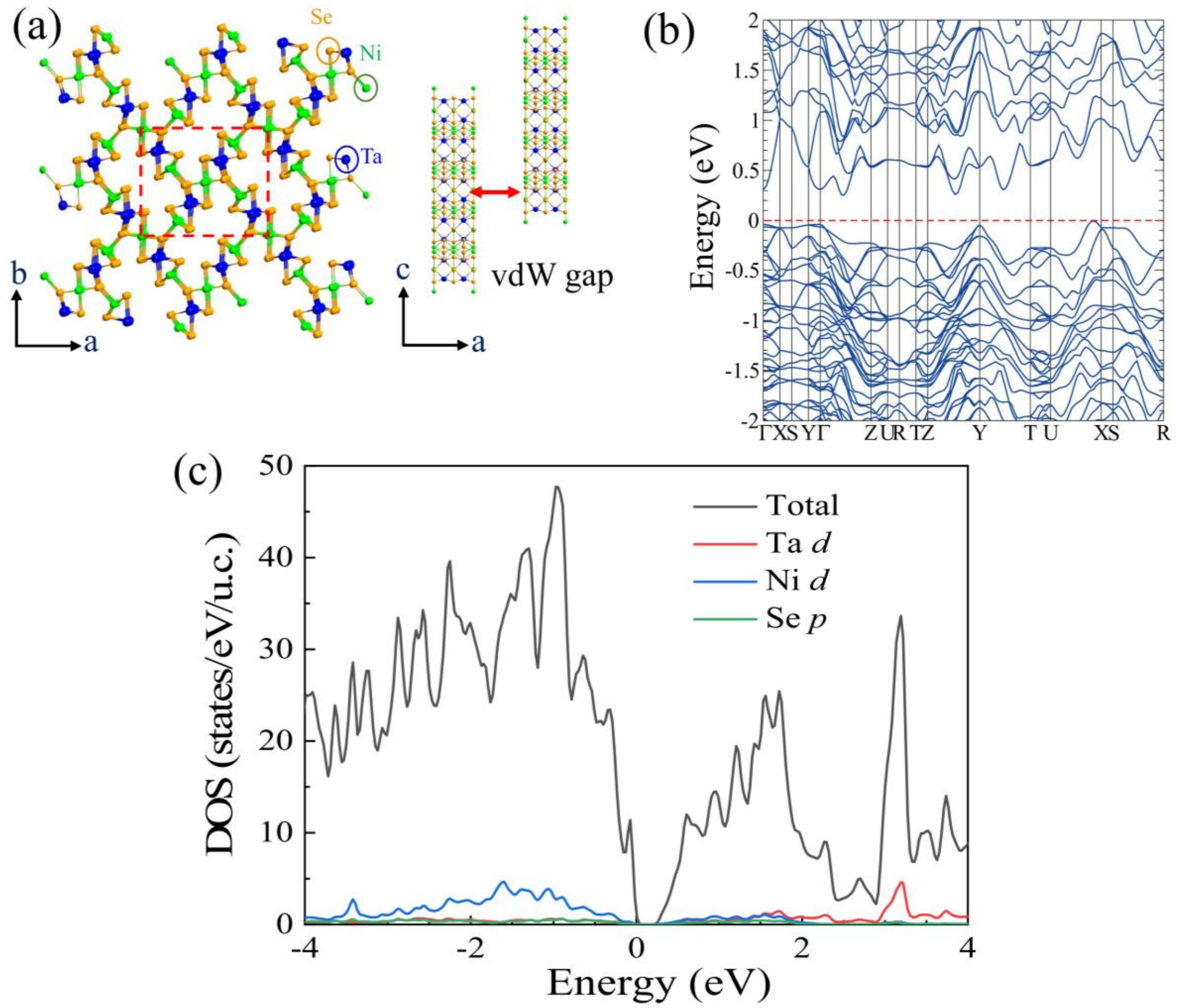

**Figure 1.** (a) Schematic crystal structure of $Ta_2Ni_3Se_8$. Blue, green and orange balls represent Ta, Ni, Se atoms, respectively. (b) Band structure along the high symmetry directions in the reciprocal space. (c) Total and partial density of states for $Ta_2Ni_3Se_8$. The Fermi energy is set at the valence band maximum.

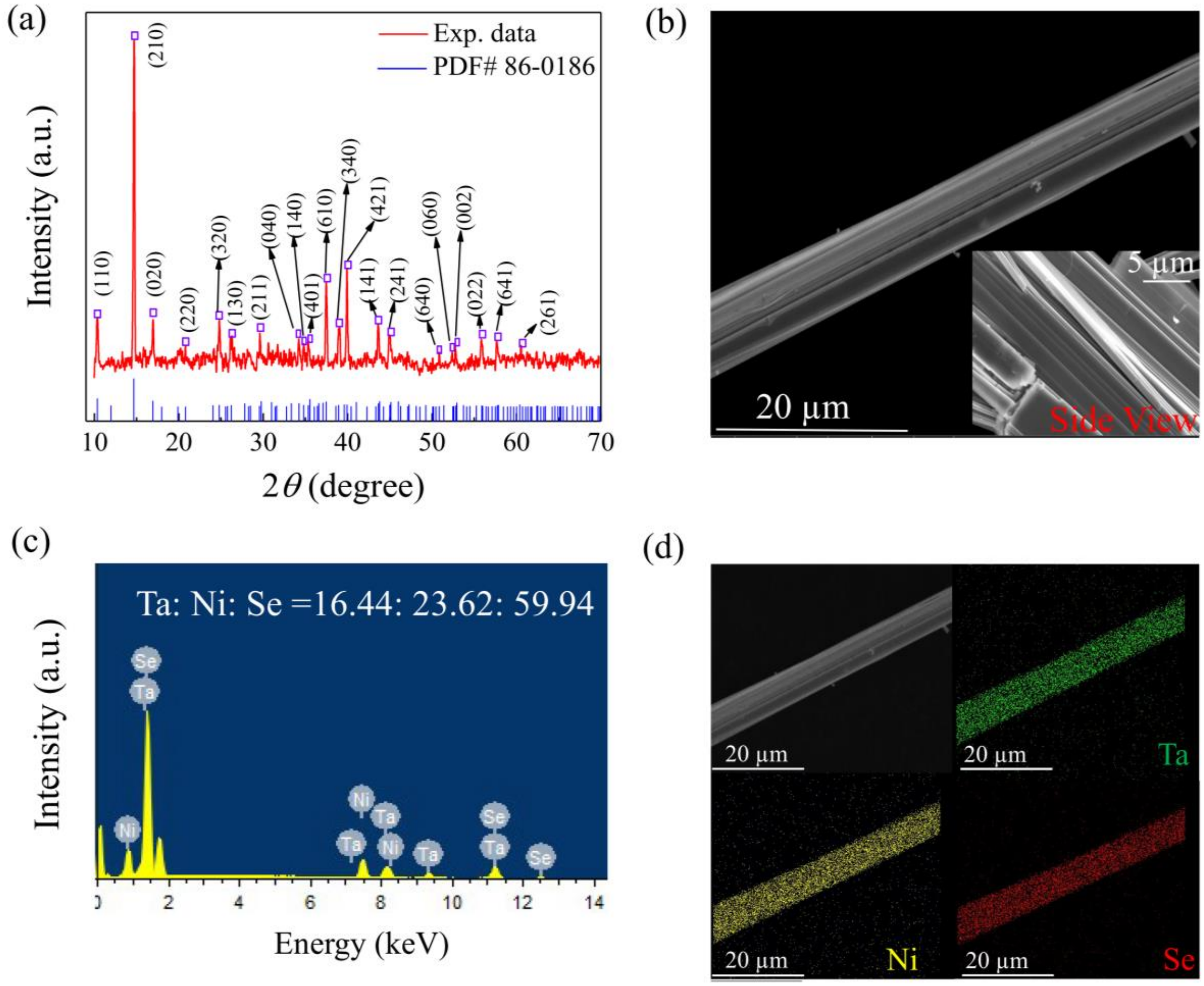

**Figure 2.** (a) XRD pattern of the as-grown $Ta_2Ni_3Se_8$ crystals. (b) SEM image of a $Ta_2Ni_3Se_8$ wire. Inset: SEM image of side view showing quasi-1D structure with layered texture. (c) EDS spectrum reveals the atomic ratio. (d) EDS elemental mapping taken on a $Ta_2Ni_3Se_8$ wire.

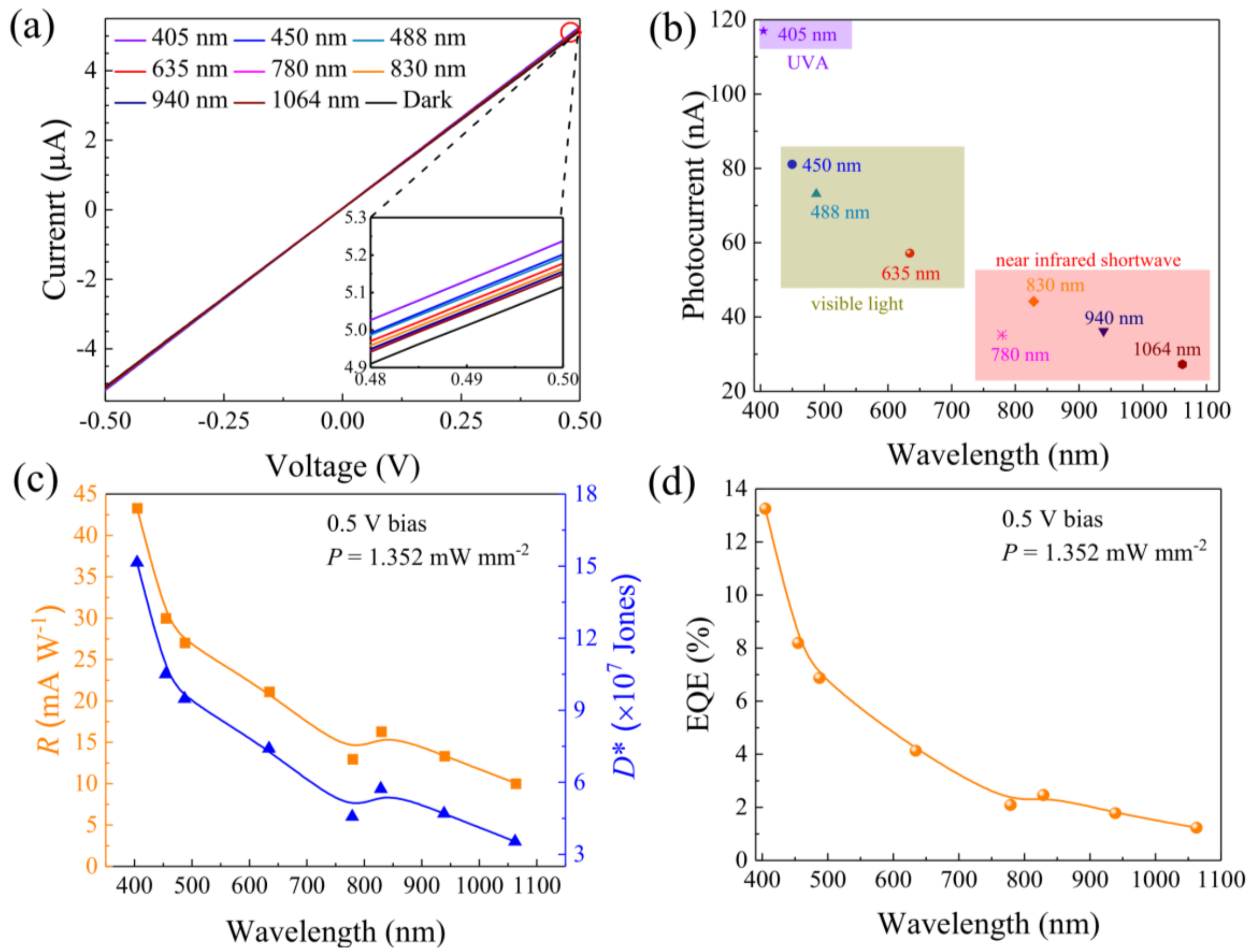

**Figure 3.** (a) *I*-*V* characteristic curves taken in the dark and under laser illumination of different wavelengths. Inset: an enlarged view of the *I*-*V* curves in the range of 0.48 V to 0.5 V. (b) Photocurrents for various wavelengths under 0.5 V bias. (c) Responsivity (left) and specific detectivity (right) and (d) EQE as a function of laser wavelength.

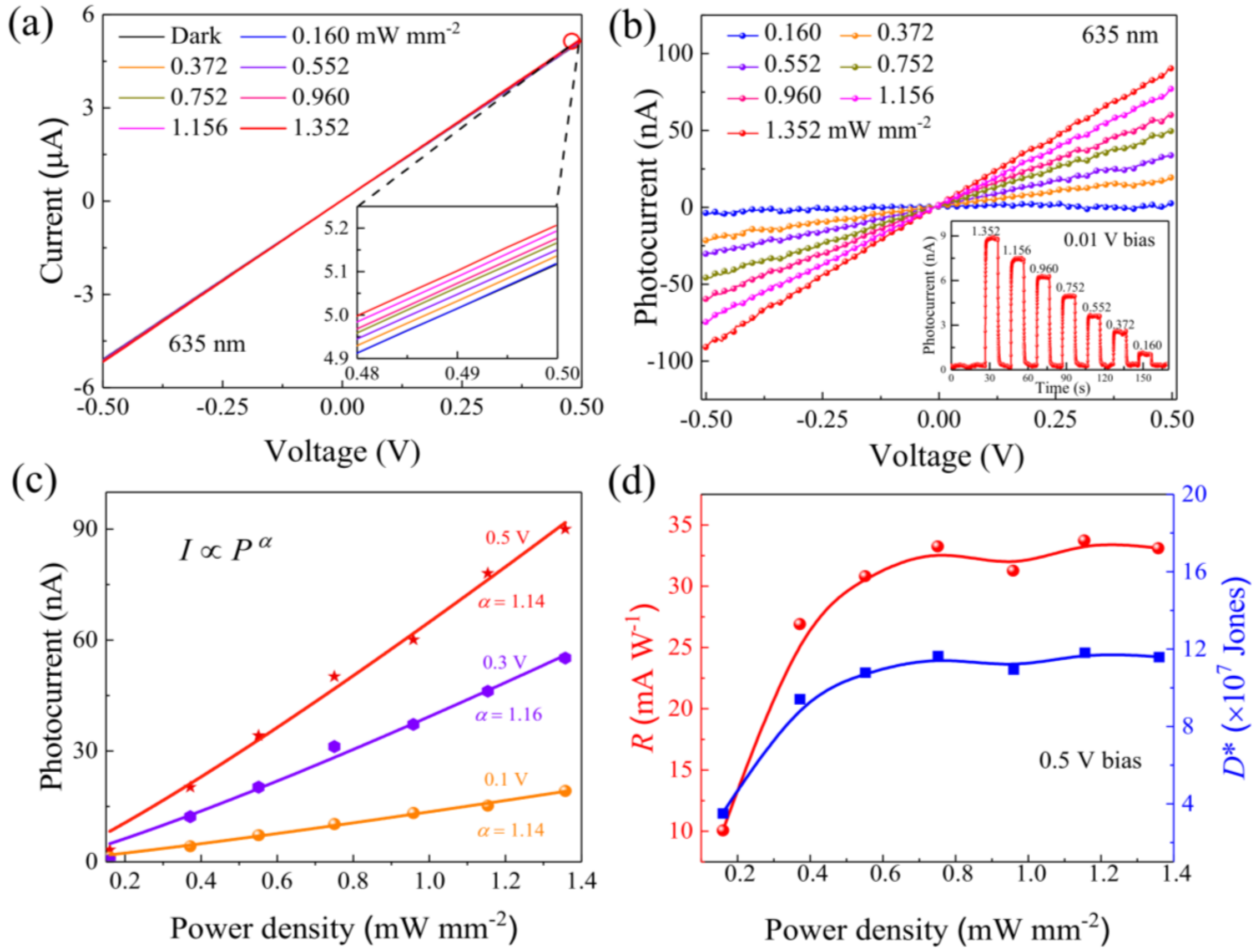

**Figure 4.** (a) *I*-*V* curves taken in the dark and under 635 nm laser tuned to different power densities. Inset: an enlarged view of the *I*-*V* curves in the range of 0.48 V to 0.5 V. (b) Photocurrent as a function of bias voltage for different laser power densities. Inset: Time-resolved photocurrent taken under different power densities and 0.01 V bias. (c) Photocurrent vs laser power density under 0.1 V, 0.3 V and 0.5 V biases. The curves represent the power-law fit. (d) Responsivity (left) and specific detectivity (right) as a function of laser power density. The curves are guides for eyes.

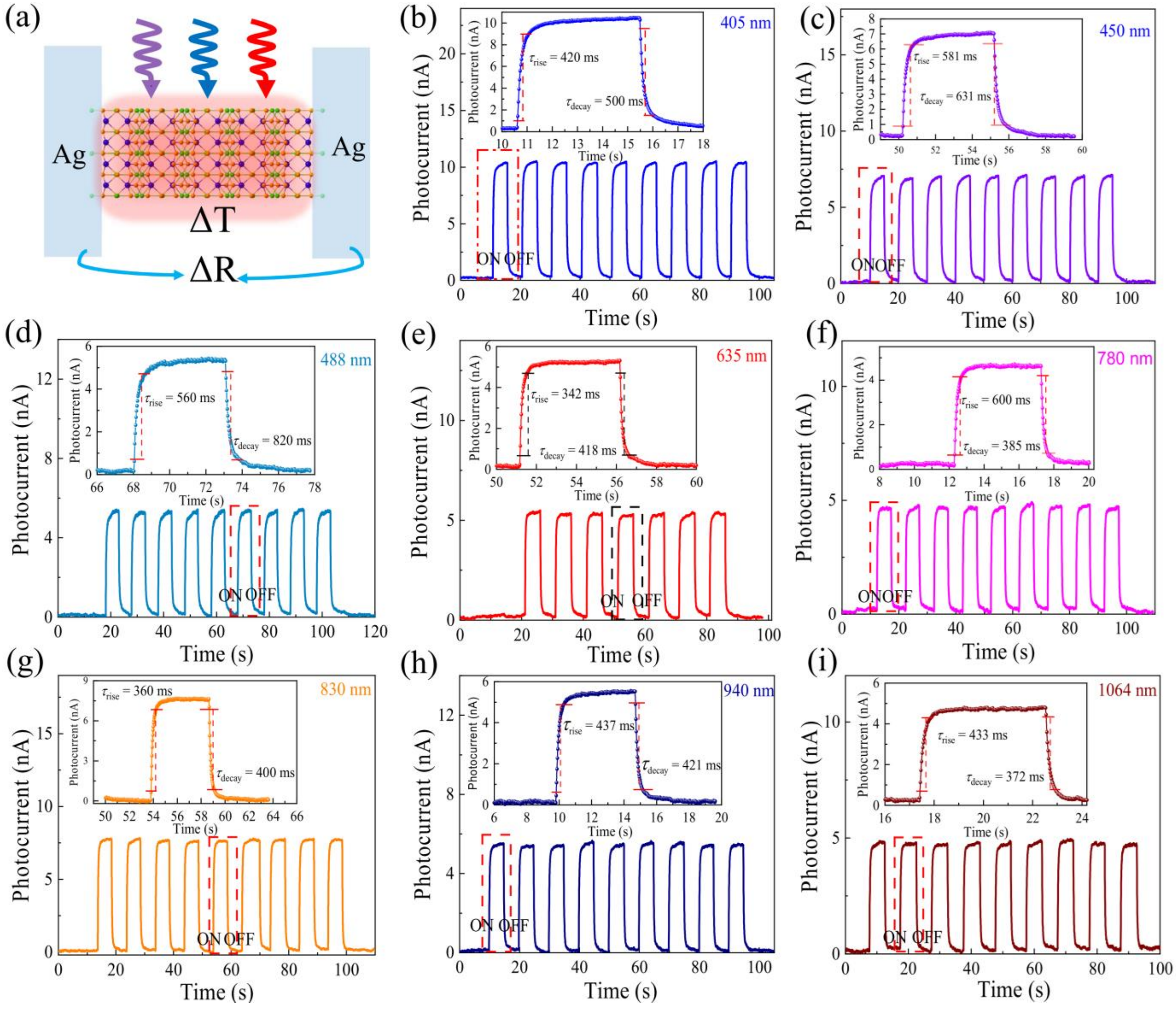

**Figure 5.** (a) Schematic illustration of the photo-bolometric effect. (b)-(i) Time-resolved photocurrent at 0.01 V bias and $P = 1.352$ mW/mm$^2$ under periodic illumination of 405 nm, 450 nm, 488 nm, 635 nm, 780 nm, 830 nm, 940 nm and 1064 nm, respectively. Insets: A single ON-OFF cycle showing determination of response time.

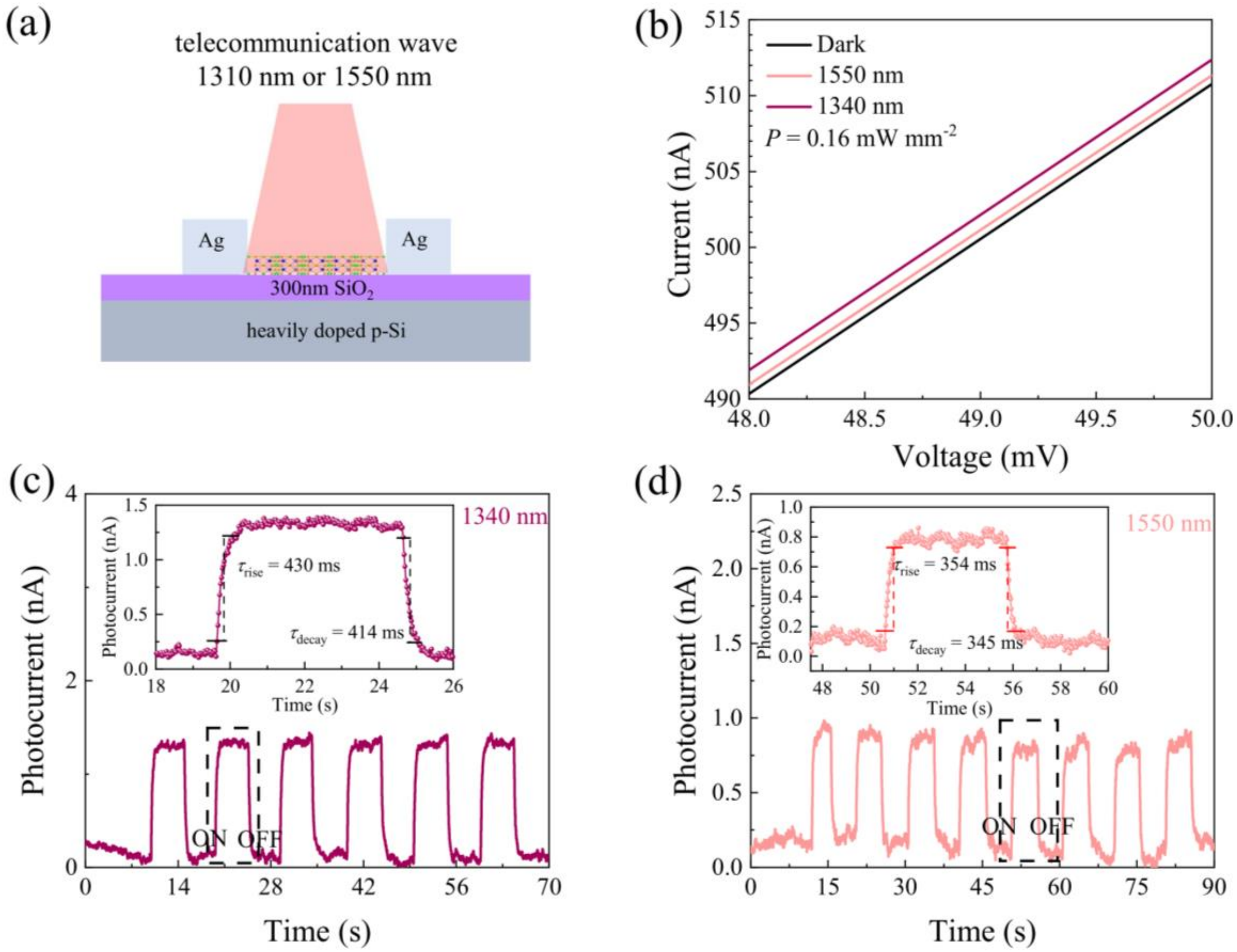

**Figure 6.** (a) Schematic illustration of $Ta_2Ni_3Se_8$ photodetector operating at 1310 nm or 1550 nm telecommunication wavelength. (b) *I-V* curves taken in the dark and at 1310 nm and 1550 nm wavelengths tuned to 0.16 mW mm$^{-2}$ in the range of 0.48 V to 0.5 V. (c) (d) Time-resolved photocurrent at 0.01 V bias and $P$ = 0.16 mW mm$^{-2}$ under periodic radiation of 1310 nm and 1550 nm, respectively. Insets: A single ON-OFF cycle showing determination of response time.

**Table 1.** Comparison of performance parameters of typical broadband photodetectors.

| Materials | Mechanism | Bias (V) | Detection range | $R$ (mA W$^{-1}$) | $D^*$ (×10$^8$ Jones) | $\tau_{rise}/\tau_{decay}$ (ms) |
|---|---|---|---|---|---|---|
| BP flake [45] | PC | 0.2 | 405-940 nm | 4.8 | - | ~1/~4 |
| $MoS_2$ flake [46] | PC | 0.1 | 455-850 nm | 50-120 | 100-1000 | >1000 |
| RGO film [47] | PB | 1.0 | UV-THz | 0.3-1.3 | - | ~135/~150 |
| $FePSe_3$ flake [48] | PC | 0.1 | 450-940 nm | 3.9 | 1.17 | ~180/~400 |
| $CsCu_2I_3$ bulk [49] | PC | 10 | 300-700 nm | 10-52 | 180-930 | ~0.2/~15 |
| $ZrS_3$ bulk [50] | PC | 1.0 | Visible light | 0.12 | 64.3 | 1000/1800 |
| $MoSe_2$ bulk [51] | PC | 1.0 | 400-700 nm | 2.35 | 0.023 | - |
| $Cs_3Bi_2I_9$ bulk [33] | PC | 1.0 | 343-4000 nm | < 0.59 | < 1 | ~4000/~10000 |
| $Ta_2Ni_3Se_8$ bulk (this work) | PB | 0.5 | 405-1550 nm | 10-43.3 | 0.35-1.51 | ~400/~400 |

Supplementary information for

# Broadband Photoresponse Arising from Photo-Bolometric Effect in Quasi-One-Dimensional $Ta_2Ni_3Se_8$

## 1. Optical image of $Ta_2Ni_3Se_8$ crystals, Raman and EDS data

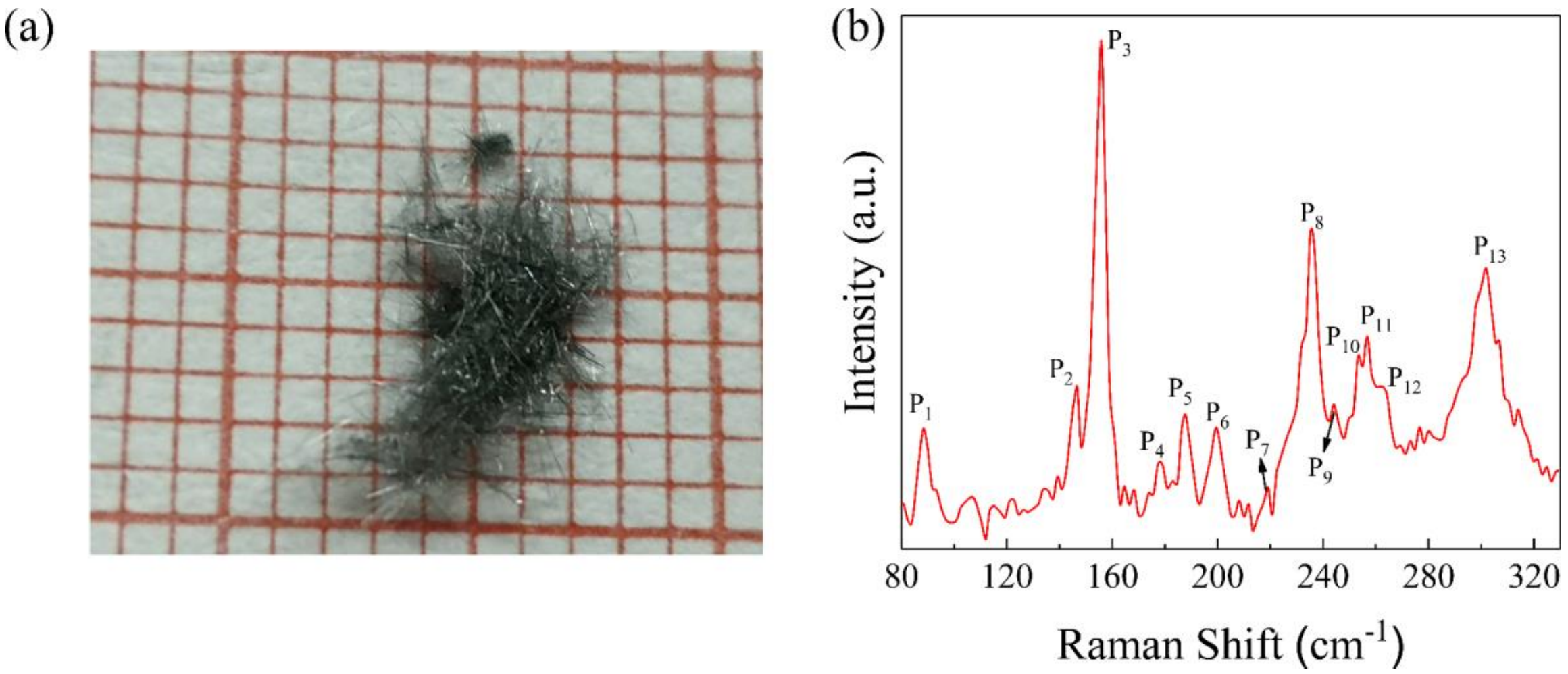

**Figure S1.** (a) An optical image of the as-grown $Ta_2Ni_3Se_8$ crystals. The size of each small grid is $1 \times 1$ $mm^2$. (b) Raman spectrum of $Ta_2Ni_3Se_8$.

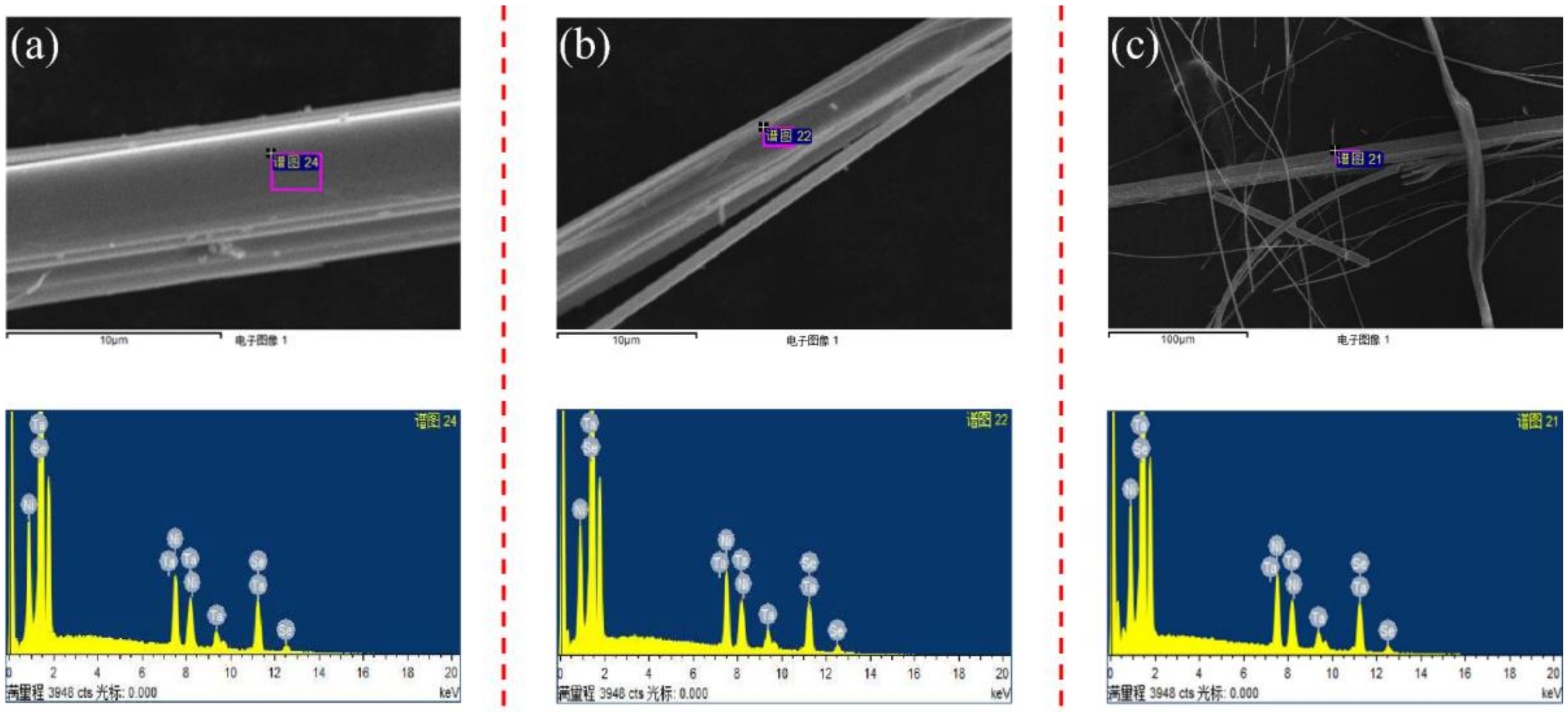

| Element | Weight% | Atomic% |
|---|---|---|
| Ta M | 33.50 | 17.01 |
| Ni K | 13.95 | 21.83 |
| Se L | 52.55 | 61.16 |
| Total | 100 | 100 |

| Element | Weight% | Atomic% |
|---|---|---|
| Ta M | 33.47 | 16.95 |
| Ni K | 14.56 | 22.73 |
| Se L | 51.97 | 60.32 |
| Total | 100 | 100 |

| Element | Weight% | Atomic% |
|---|---|---|
| Ta M | 33.88 | 17.90 |
| Ni K | 13.57 | 21.47 |
| Se L | 51.55 | 60.63 |
| Total | 100 | 100 |

**Figure S2.** EDS data of different $Ta_2Ni_3Se_8$ crystals.

## 2. Optoelectronic measurements of devices D2 and D3

The optoelectronic properties of two additional devices, namely D2 and D3, were measured. The channel dimensions of D2 and D3 are approximately 1.5 μm (width) × 2.4 mm (length) and 1.7 μm (width) × 3 mm (length), respectively. Compared with the first device D1 with dimensions of 1 μm (width) × 2 mm (length), these two devices reveal no significant differences. All of these devices exhibit a broadband photoresponse. The $\alpha$ values obtained from the power-law fits are also greater than 1, which indicates that the photocurrent generation mechanism of D2 and D3 is also the photo-bolometric effect.

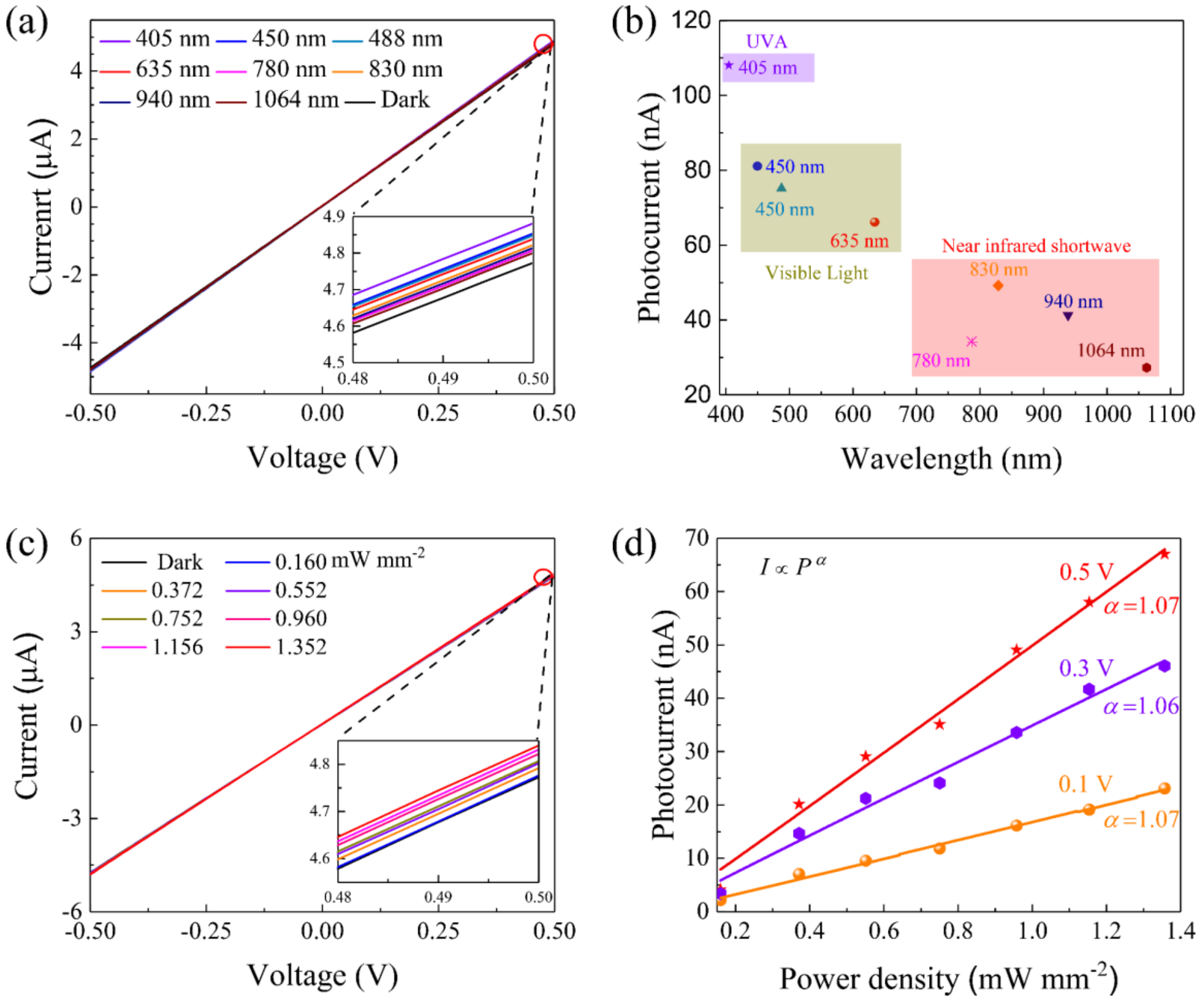

**Figure S3.** Optoelectronic measurements of device D2. (a) *I-V* characteristic curves taken in the dark and under laser illumination of different wavelengths. Inset: an enlarged view of the *I-V* curves in the range of 0.48 V to 0.5 V. (b) Photocurrents for various wavelengths under 0.5 V bias. (c) *I-V* curves taken in the dark and under 635 nm laser tuned to different power densities. Inset: an enlarged view of the *I-V* curves in the range of 0.48 V to 0.5 V. (d) Photocurrent vs laser power density under 0.1 V, 0.3V and 0.5 V biases. The curves represent the power-law fit.

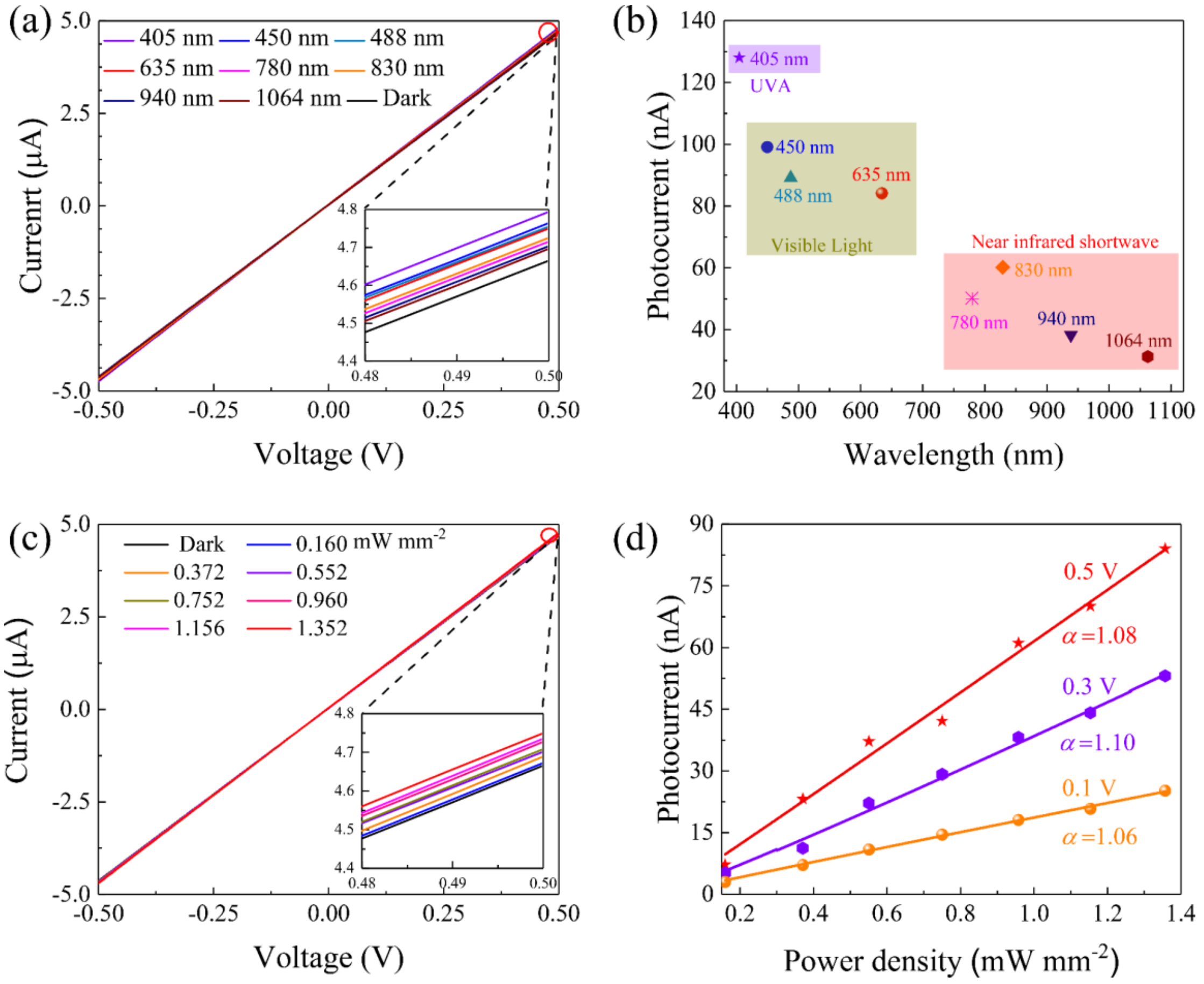

**Figure S4.** Optoelectronic measurements of device D3. (a) *I-V* characteristic curves taken in the dark and under laser illumination of different wavelengths. Inset: an enlarged view of the *I-V* curves in the range of 0.48 V to 0.5 V. (b) Photocurrents for various wavelengths under 0.5 V bias. (c) *I-V* curves taken in the dark and under 635 nm laser tuned to different power densities. Inset: an enlarged view of the *I-V* curves in the range of 0.48 V to 0.5 V. (d) Photocurrent vs laser power density under 0.1 V, 0.3V and 0.5 V biases. The curves represent the power-law fit.

## 3. Long-term stability

Additional experiments were performed to test the long-term stability. First, the device operated for 30 response cycles with repeated illumination on and off. In each cycle, the illumination on and off time was set to 10 s. The power density was set to a high value, i.e., 1.352 mW mm$^{-2}$. As shown in figure S5(a), the device exhibits high stability and repeatability. Second, the device was tested under long time illumination (200 s). We can see that the stability of the device is also good enough in the long-term ON state. Once the illumination is turned off, the device can quickly return to the dark state, and the dark state is also stable after long-term operation. Combining the two tests, it is affirmed that the device has very good long-term stability and durability.

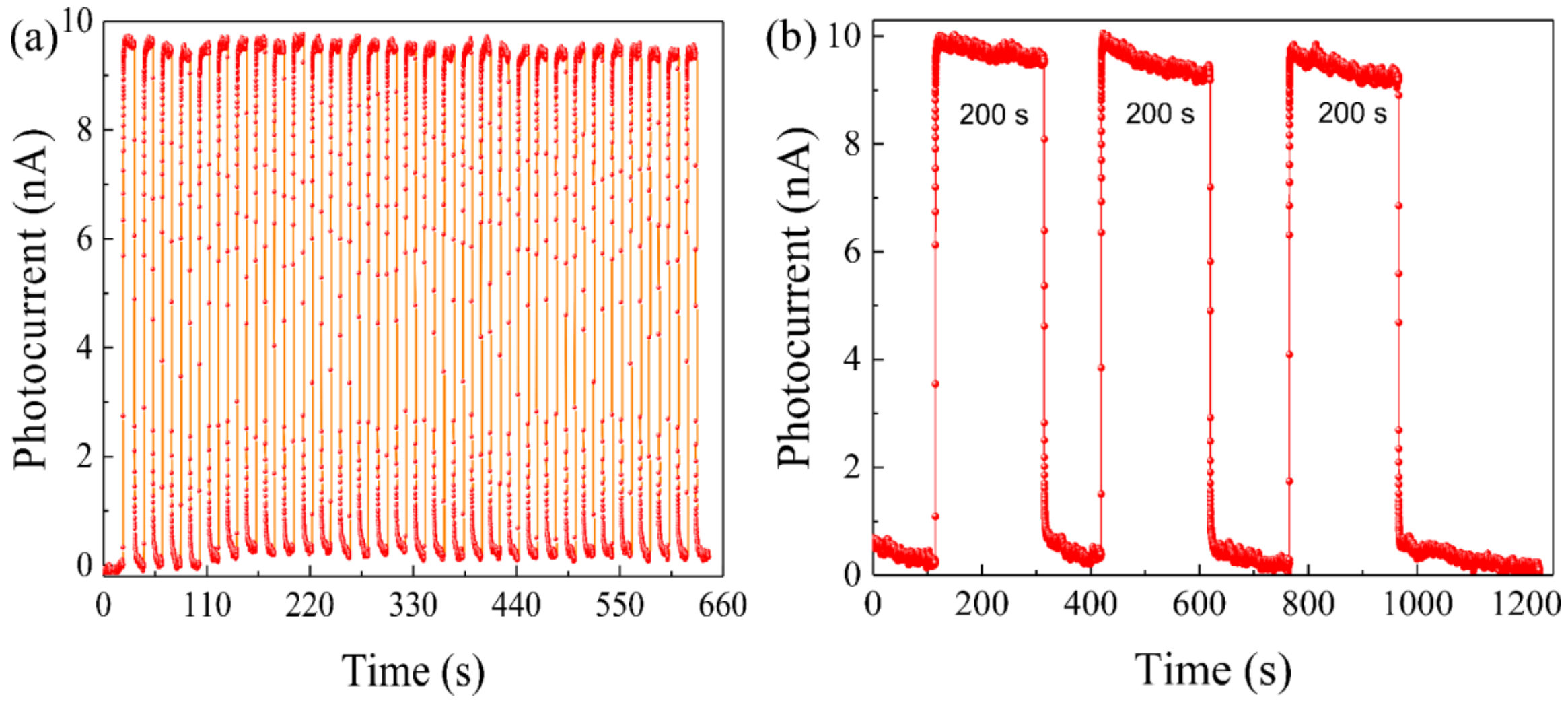

**Figure S5.** (a) 30 response cycles setting the power density to 1.352 mW $mm^{-2}$. (b) Stability of long illumination time (200 s). $\lambda$ = 635 nm and $V_{ds}$ = 0.01V.

## 4. RC constant and response time

The relationship between the RC constant and the photocurrent-time waveform is illustrated in figure S6. The response time of the device is limited by the RC constant, which usually occurs when the device is illuminated by high-frequency incident light, and the response time cannot keep up with the frequency of the light, resulting in distorted output waveform, as shown in figure S6(c). In this case, a large parasitic capacitance exists between the channel material and the contact electrodes [B. E. A. Saleh and M. C. Teich, *Fundamentals of Photonics*. (John Wiley & Sons, Inc., New York, Chichester, Brisbane, Toronto, Singapore, 1991).]. As seen in figures 3(a) and 4(a), the *I-V* curves of our devices show very good linearity, indicating that an ohmic contact is formed and the capacitance is very small. Furthermore, in our experiments, the laser illumination is repeatedly turned on and off at a low frequency of 0.05 Hz (duty ratio: 50%). Figures 5(b)-5(i) show good switching cycle stability for all curves, indicating that the photodetector can work well and durably at various wavelengths. The slowest response appears at 488 nm, i.e., $\tau_{\text{rise}}$ = 560 ms and $\tau_{\text{decay}}$ = 820 ms, both of which are much shorter than the period of the incident light (20 s). The output waveforms are undistorted, similar to that in figure S6(b). These good square waveforms correspond to small parasitic capacitances, which is consistent with the linear *I-V* curves. That is, our devices are not RC constant limited for the given parameters.

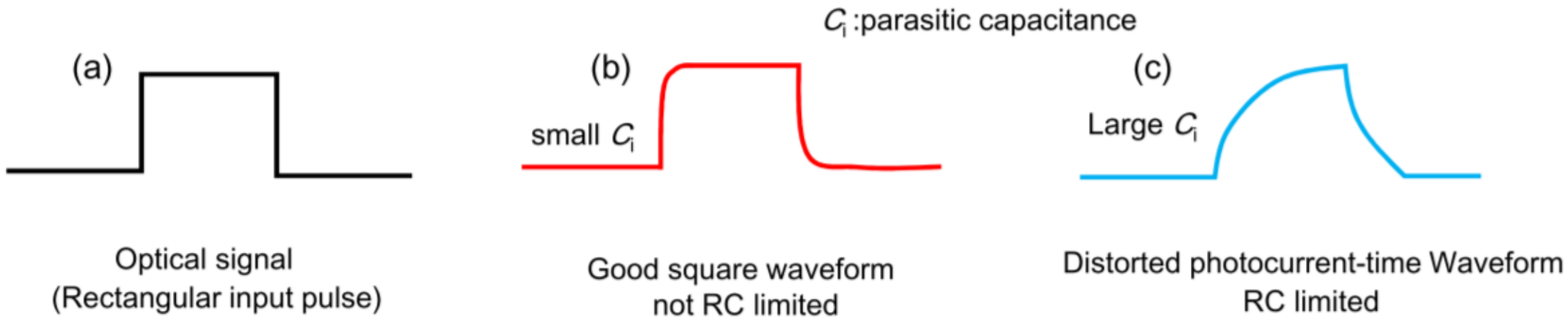

**Figure S6.** (a) Optical input signal, (b) output waveform not limited by RC constant and (c) output waveform limited by RC constant.

**5. Power density dependence of response time**

The power density dependence of the response time was investigated. As shown in figure S7, the response time is $\tau_{\text{rise}}$ = 342 ms and $\tau_{\text{decay}}$ = 418 ms for $P$ = 1.352 mW mm$^{-2}$, $\tau_{\text{rise}}$ = 320 ms and $\tau_{\text{decay}}$ = 360 ms for $P$ = 0.75 mW mm$^{-2}$, and $\tau_{\text{rise}}$ = 216 ms and $\tau_{\text{decay}}$ = 327 ms for $P$ = 1.352 mW mm$^{-2}$, respectively. It seems that the response becomes slower at higher power densities. Here is a possible explanation. Given that the response time is not limited by the RC constant, another factor limiting the response time is the carrier transit time $\tau_{\text{tr}}$, which is defined as the time the carriers take to go through the channel. It can be expressed as $\tau_{\text{tr}}=L^2/2\mu V_{\text{bias}}$, where $L$ is the channel length and $\mu$ is the carrier mobility. As $\mu$ decreases, $\tau_{\text{tr}}$ and response time increase. For photodetectors conforming to the photo-bolometric effect, the heating process caused by photon absorption affects the carrier mobility. For higher power densities, more light absorption means increased temperature and enhanced phonon scattering, which will suppress the carrier mobility. Therefore, higher power density corresponds to slower response.

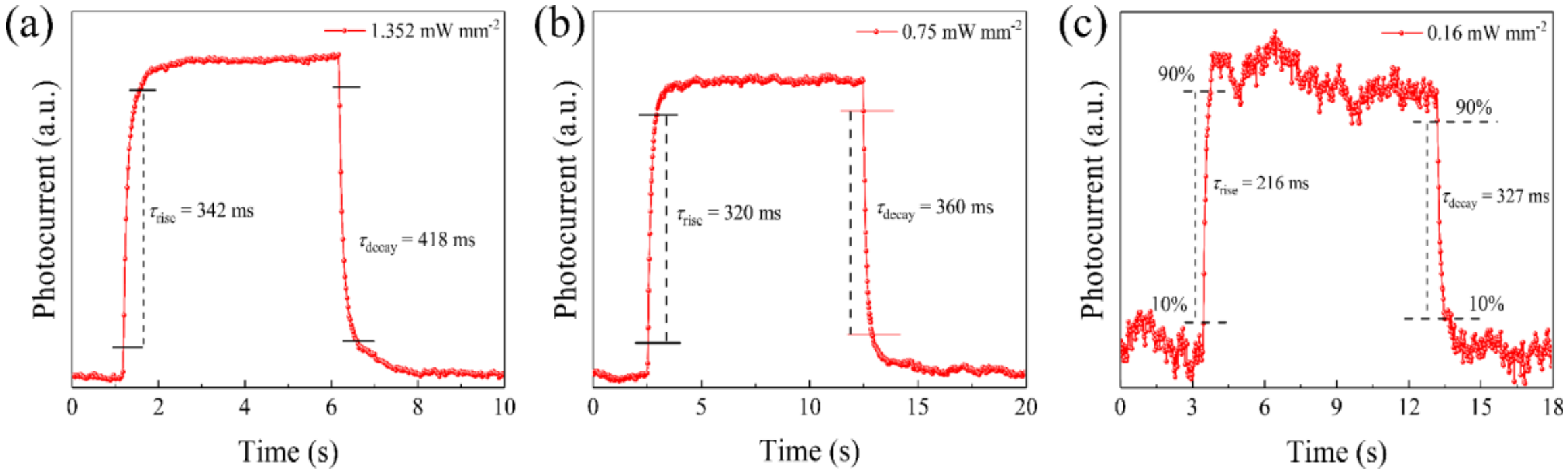

**Figure S7.** Response time measured under different power densities.